\documentclass{amsart}
\usepackage{amsmath,graphicx}
\usepackage{amssymb}
\usepackage[normalem]{ulem}
\usepackage{color}

\begin{document}

\newcommand\re[1]{(\ref{#1})}
\newcommand{\pd}[2]{\frac{\partial #1}{\partial #2}}
\newcommand{\pdt}[3]{\frac{\partial^2 #1}{\partial #2 \partial #3}}
\newcommand{\mg}[1]{\partial_{#1}}
\newcommand{\F}[2]{F^{#1}_{\;#2}}
\renewcommand{\labelitemi}{--}

\title{Theories and heat pulse experiments of non-Fourier heat conduction}
\author{P. V\'an}
\address{Department of Theoretical Physics\\
         Wigner RCP, RMKI, Budapest, Hungary and \\
Department of Energy Engineering\\
        Budapest University of Technology and Economics, Hungary\\
Montavid Thermodynamic Research Group}

\pagestyle{plain}
\markright{A short non-Fourier review}

\date{\today}

\begin{abstract}
The experimental basis and theoretical background of non-Fourier heat 
conduction is shortly reviewed from the point of view of non-equilibrium 
thermodynamics. The performance of different theories is compared in case of 
heat pulse experiments.
\end{abstract}

\maketitle

\section{Introduction}

There are several extensive reviews of heat conduction beyond Fourier 
\cite{JosPre89a,JosPre90a,Cha98a,Str11b,Cim09a,Leb14a}. When non-equilibrium 
thermodynamics is concerned, usually Extended Thermodynamics 
is in the focus of surveys in this subject. In this short paper heat conduction 
phenomena is reviewed from a broader perspective of non-equilibrium 
thermodynamics. The performance of different theories is compared on the 
example of heat pulse experiments. 

There are two preliminary remarks to underline the particular point of view of 
this work. The first remark concerns the so called heat conduction paradox 
of the infinite speed of signal propagation in Fourier theory. Infinite 
propagation speeds are frequently mentioned as a disadvantage 
\cite{Cat48a,JouAta92b} and this is a main motivation for looking 
symmetric hyperbolic evolution equations in general \cite{MulRug98b}. However
\begin{itemize}
\item the relativistic formulation of heat conduction is a different matter and 
parabolic equations may be compatible with relativity 
\cite{KosLiu00a,HerPav01a,KosLiu01a}. 
\item The finite characteristic speeds of hyperbolic equations are determined 
by material parameters. Therefore, in principle, in nonrelativistic spacetime 
they can be larger than the speed of light.
\item In case of parabolic equations the speed of real signal propagation is 
finite, e.g. because the validity condition of a continuum theory determines a 
propagation speed. Moreover, the sensitivity of the experimental devices is 
finite, therefore they can measure only finite speeds 
\cite{Fic92a,KosLiu00a,Cim04a,VanBir08a}.
\end{itemize}  

Our second remark is about the role of non-equilibrium thermodynamics as a 
frame theory of several physical disciplines. Here the fundamental assumption 
is the validity of the second law of thermodynamics in the form of an entropy 
balance with nonnegative production. If  microscopic or 
mesoscopic theories do not want to violate macroscopic second law, they must 
suit the requirements derived from the phenomenological conditions. This is the 
extension of the classical universality of the absolute temperature. The 
validity of the second law is a weak assumption from a physical point of view 
but introduces remarkably strict restrictions for the evolution equations. 

According to these arguments it is no unreasonable to look for parabolic or at 
least non-hyperbolic extensions of the Fourier equation of heat conduction that 
fulfill basic principles of continuum physics.

In the following section we survey the most important experimental observations 
of non-Fourier heat conduction. Then the relevant kinetic and thermodynamic  
theories are surveyed. Their assumptions, properties and modeling capabilities 
are compared on the example of heat pulse experiments.

\section{Experiments}

\subsection{Helium II}
The first experimental observation of non-Fourier heat conduction was the 
measurement of wave like propagation of heat in liquid Helium II by Peshkov 
\cite{Pes44a}. The experiment was motivated by two-fluid theories of Tisza 
\cite{Tisz38a} and Landau \cite{Lan41a}. Landau suggested that wave like 
propagation is a property of phonon gas and introduced the terminology "second 
sound". The works of Cattaneo, Morse-Feshbach and Vernotte 
\cite{Cat48a,MorFes53b,Ver58a} indicated that the related phenomena may be more 
general, there is an "inertia of heat".

\subsection{Low temperature solids}

Phonon gas theory predicted the existence of second sound in solids. The 
key aspects of the observation were the modeling of phonon scattering by 
kinetic theory \cite{GuyKru64a,GuyKru66a1,GuyKru66a2} and the application of 
heat pulses. The derivation of a dissipative extension of the 
Maxwell-Cattaneo-Vernotte equation from kinetic theory, the 
Guyer-Krumhansl equation \cite{GuyKru66a1}, provided the "window 
condition", when dissipation is minimal. A careful preparation of 
crystals with particular properties resulted in observation of second sound in 
solid $^3$He, $^4$He, NaF an Bi crystals  
\cite{AckGuy68a,AckOve69a,McNEta70a,NarDyn72a}. In 
some experiments second sound and ballistic propagation -- heat pulses 
propagating with the speed of sound -- were observed together \cite{JacWal71a}. 
At lower temperatures ballistic propagation appears without second sound. 

The correct modeling of the observed parallel ballistic, wave like and 
diffusive propagation of heat in a uniform theoretical framework is the most 
important benchmark of the theories.

\subsection{Heterogeneous materials at room temperature}

The low temperature heat pulse measurements in dielectric crystals exploit well 
understood microscopic mechanisms. At room temperature various dissipative 
processes suppress these effects. However, some experiments 
with heterogeneous materials indicated the possibility of non-Fourier heat 
conduction at room temperature, too \cite{Kam90a,MitEta95a}. These experiments 
were not confirmed, more properly the attempts of exact reproduction of these 
experiments are contradictory \cite{HerBec00a,HerBec00a1,RoeEta03a,ScoEta09a}.

\subsection{Small size} 

Nano heat conduction is a popular research field with interesting new 
observations \cite{DasEta03a}. Non-Fourier effects are in principle enhanced by 
reducing the size of the samples and the speed of the phenomena 
\cite{KimEta01a,CahEta03a,Zha07b}.  Guyer-Krumhansl equation is extensively 
analyzed in this respect \cite{CimEta09a,CimEta13a}. 

\subsection{Relativistic experiments}

Relativistic fluid effects became experimentally available in quark-gluon 
plasma. Experiments in RHIC and LHC confirmed the existence of a dissipative 
relativistic fluid \cite{LuzRom08a}. In relativistic theories the inertial 
effects cannot be neglected, viable theories incorporate wave like heat 
propagation, due to stability problems \cite{Mul67a,Isr76a,IsrSte76a}. However, 
in case of ultrarelativistic speeds 
the flow is energy dominated, therefore it is reasonable to use Landau-Lifshitz 
flow-frame, where propagation of energy defines the flow. This choice 
eliminates the possibility of heat conduction 
and only viscous dissipation is possible. According to thermodynamic arguments, 
the  choice of flow-frames is not arbitrary and in case of lower speeds heat 
conduction may be separated from viscous dissipation 
\cite{VanBir12a}.

\section{Theories}

The three main approaches are kinetic theory, non-equilibrium thermodynamics 
and continuum mechanics combined with non-Fourier heat conduction. 
Up to now the relation of these general approaches is not clear. Fortunately 
these general frame theories offer several possibilities of comparison, 
validation and extension far  beyond heat conduction phenomena. In particular 
we will see, that the performance of these theories in modeling ballistic and 
diffusive heat propagation is different. 

\section{Kinetic theory}

As it was already mentioned, kinetic theory played an important predictive role 
in understanding and designing proper experiments with the help of 
identified microscopic heat propagation mechanisms. In particular the relation 
of pure kinetic and continuum approaches is important in this respect. 
Kinetic theory introduces a hierarchical structure of macroscopic field 
quantities with coupled balance form evolution equations. In this balance 
structure current density at a given level is a source density at the next 
level. The tensorial order of the $n$-th variable 
is $n$. For phonons  with the Callaway collision integral one may obtain the 
following system of equations in one spatial dimension (\cite{MulRug98b} p349):

\begin{equation}
\partial_t u_n + 
\frac{n^2}{4n^2-1}c\partial_x u_{n-1} + c \partial_x u_{n+1} =\left \{ 
\begin{array}{ll}
\displaystyle
0 \ & \ n=0, \\
-\frac{1}{\tau_R}u_1 \ & \ n=1, \\
- \left( \frac{1}{\tau_R}+\frac{1}{\tau_N} \right )u_{n} \ & \ 
n\geq 2.
\end{array} \right.
\end{equation}

Here $\partial_t$ and $\partial_x$ are the time and space derivatives, $u_n$ is 
related to the $n$th momentum of the one particle probability 
distribution function by constant multipliers, $\tau_R$ and $\tau_N$ are 
relaxation times, $c$ is the Debye speed. A truncated hierarchy at the third 
moment, $u_n = 0$ of $n\geq 3$, results in the following set of equations:
\begin{eqnarray}
\partial_t u_0 + c\partial_x u_{1} &=& 0,\\
\partial_t u_1 + 
\frac{1}{3}c\partial_x u_{0} + c\partial_x u_{2} &=& -\frac{1}{\tau_R}u_1,\\
\partial_t u_2 + \frac{4}{15}c\partial_x u_{1}   &=& -\left( 
\frac{1}{\tau_R}+\frac{1}{\tau_N} \right)u_{2}.
\end{eqnarray}

The macroscopic fields are introduced by the following definitions: 
the energy density $e=\hbar c u_0$, heat flux $q=\hbar c^2 u_1$ and the moment 
of the heat flux $Q=\hbar c^2 u_1$. One may apply the (approximate) caloric 
equation of state $e= \rho\hat c T$ with constant density $\rho$ 
and specific heat $\hat c$. Moreover, it is convenient to redefine the 
coefficients by introducing the $q$-relaxation time $\tau_q=\tau_R$, the 
$Q$-relaxation time $\tau_Q= \frac{\tau_R\tau_N}{\tau_R+\tau_N}$, the Fourier 
heat conduction coefficient $\lambda= \rho \hat c \tau_R c^3/3$ and two 
additional phenomenological coefficients $k_{21} = \tau_R c^2$ and $k_{12} = 4 
\tau/15$. Then the previous equations may be written as:
 \begin{eqnarray}
 \partial_t e + \partial_x q &=& 0,\label{ebal}\\
 \tau_q\partial_t q + \lambda \partial_x T + \kappa_{21}\partial_x Q &=& -q, 
 \label{qbal}\\
 \tau_Q\partial_t Q + \kappa_{12}\partial_x q &=& - Q \label{Qbal}.
 \end{eqnarray}

If $\tau_Q=0$, one may obtain the Guyer-Krumhansl equation eliminating $Q$ from 
\re{qbal}-\re{Qbal}. If  $\kappa_{21}=0$ then \re{qbal} is the 
Maxwell-Cattaneo-Vernotte equation.

In Rational Extended Thermodynamics (RET) the coefficients of the 
above set of field equations are calculated exactly as it was shown above. 
Remarkable that one needs only three parameters, 
the Debye speed and the R and N process related relaxation times 
\cite{MulRug98b,CimEta14a,Leb14a}. 

Extended Irreversible Thermodynamics (EIT) is a related but different approach, 
where the structure of the equations is important, but the values of the 
coefficients 
are considered mostly undetermined, therefore e.g. in \re{ebal}-\re{Qbal} the 
number of parameters is higher \cite{JouAta92b,LebEta08b,CimEta14a}. In RET the 
microstructure is unavoidable in EIT the microstructure is flexible.  In RET 
the coefficients are to be calculated from a microscopic model, in EIT  
some coefficients are to be measured. RET is a local theory by construction, 
EIT may be weakly nonlocal. RET is strictly compatible with kinetic theory, EIT 
is weakly compatible with kinetic theory. In EIT the derivation of the 
Guyer-Krumhansl equation either refers to the momentum hierarchy 
\cite{LebEta08b} or weakly nonlocal extensions in a pure non-equilibrium 
thermodynamic framework \cite{LebAta98a,CimEta13a}.

The experimental results of heat conduction of mixed ballistic-wavy-diffusive 
propagation can be modeled in this framework. However, with the restriction of 
using only 3 parameters RET requires about 30 (!) moments in order to obtain 
correct propagation speeds both for ballistic heat propagation and second sound 
\cite{DreStr93a,MulRug98b}. In EIT, considering some coefficients as 
phenomenological parameters, in principle it is possible to incorporate correct 
propagation speeds for the second sound and also for the ballistic phonons with 
the above set of equations. However, in this case one cannot stop at the level 
of the Guyer-Krumhansl equation, and, most importantly, more than 3 parameters 
are required. A hierarchical phenomenological background theory is necessary to 
justify these assumptions. 


\section{Non-equilibrium thermodynamics}

Irreversible thermodynamics as field theory was established by Eckart, with the 
assumption of local equilibrium \cite{Eck40a1,Eck40a2,Eck40a3,Eck48a}. 
Maxwell-Cattaneo-Vernotte type extensions require a deviation from 
local equilibrium. In this paper {\em non-equilibrium thermodynamics} is a 
collective nomination of thermodynamic theories based on the direct 
exploitation of the entropy inequality with or without the requirement of local 
equilibrium. Classical Irreversible Thermodynamics, Extended Thermodynamics, 
Thermodynamics with Internal Variables are special theories of non-equilibrium 
thermodynamics \cite{CimEta14a}. 

The first basic assumption extending non-equilibrium thermodynamics 
beyond local equilibrium is that the thermodynamic fluxes of Classical 
Irreversible Thermodynamics are state variables. Then Maxwell-Cattaneo-Vernotte 
equation is a consequence of the second law by thermodynamical flux-force 
linearization. This 
idea was first proposed by M\"uller \cite{Mul67a1}. One of M\"uller's argument 
was the compatibility with kinetic theory. An other independent way is based on 
the seminal treatment of discrete systems by Machlup and Onsager, who 
introduced kinetic energy of thermodynamic state variables \cite{MacOns53a}. 
The continuum generalization of Gyarmati is based on variational considerations 
and therefore introduced thermodynamic fluxes as independent variables instead 
of time derivatives \cite{Gya77a}. Later on this idea was further generalized 
by introducing arbitrary internal variables for the same purpose 
\cite{Ver85b,Ver97b}. 

As it was mentioned previously, thermodynamic fluxes as state variables may 
lead to the Maxwell-Cattaneo-Vernotte equation, but to obtain the 
Guyer-Krumhansl equation requires further assumptions. The modification of the 
entropy current is a straightforward idea in this respect. The idea, that the 
entropy current density is not always the heat flux divided by the temperature 
is natural in mixtures and also in Extended Thermodynamics \cite{Mul67a}, but 
may be natural in other generalizations, too 
\cite{Ver83a,CimVan05a,CiaAta07a}. 
To obtain the non-equilibrium thermodynamic counterpart of \re{qbal}-\re{Qbal} 
it is reasonable to assume that the additional fields are the heat flux vector 
${q}$ and a second order tensorial internal variable ${Q}$ and use the Ny\'iri 
form generalization of the entropy current density \cite{Nyi91a1,Van01a2}.

In this paper we restrict ourselves to rigid heat conductors in a one 
dimensional treatment, therefore the time derivatives are partial and the space 
derivatives are one directional. For a more complete treatment see 
\cite{KovVan15a}. Our starting point is the balance of entropy, which results 
in nonnegative production when constrained by the balance of  internal energy 
\re{ebal}:
\begin{equation}
\partial_t s + \partial_x J \geq 0.
\label{s_bal}\end{equation}
Here $s$ is the entropy density and $J$ is the entropy current density. 

Deviation from local equilibrium will be characterized by two basic 
constitutive hypotheses:
\begin{itemize}
\item We assume that the entropy density depends quadratically on the 
additional fields:
\begin{equation}
s(e, q, Q) = s_{eq}(e) - \frac{m_1}{2} q^2 - \frac{m_2}{2} Q^2,
\label{neqs}\end{equation}
where $m_1$ and $m_2$ are constant, nonnegative material coefficients. The 
derivative of the local equilibrium part of the entropy function 
$s_{eq}$ by the internal energy is the reciprocal temperature: 
$$\frac{d s_{eq}}{d e} = \frac{1}{T}.$$
The quadratic 
form is introduced for the sake of simplicity and may be considered as a first 
approximation. The coefficients $m_1$ and $m_2$ are nonnegative 
because of the concavity of the entropy function, that is, thermodynamic 
stability.
\item We assume that the entropy flux is zero if $q=0$ and $Q=0$. Therefore 
it can be written in the following form:
\begin{equation}
J =  b q+ B  Q.
\label{neqJ}\end{equation}
Here $b$ is originated from a second order tensorial constitutive function and 
$B$ from a third order one. They are the current multipliers \cite{Nyi91a1}. 
\end{itemize}

Here the basic fields are $T,q$ and $Q$, the constitutive 
functions are $b$ and $B$. The entropy production can be calculated 
accordingly:
\begin{eqnarray}
\partial_t s+ \partial_x J 
&=&	-\frac{1}{T}\partial_x q - 	m_1 q\partial_t q -	m_2 Q \partial_t Q + 
	b\partial_x q + \nonumber\\
& &	q\partial_x b +	B\partial_x Q + Q\partial_x B \nonumber\\
&=& \left(b - \frac{1}{T}\right) q + 
	\left(\partial_x b -  m_1 \partial_t q\right)q -\left(\partial_xB -  m_2 
	\partial_t Q\right) Q +	B \partial_x Q \geq 0. 
\label{entrpr}\end{eqnarray}
In the last row the first and third terms are products of second order tensors, 
the second term is a product of vectors and the last term is of third order 
ones. It is not apparent in our one dimensional simple notation. The time 
derivatives of the state variables $q$ and $Q$ represent their 
evolution equations, which are considered as constitutive quantities together 
with the current multipliers $b$ and  $B$.
Therefore, one can identify four thermodynamic forces and currents in the above 
expression and assume linear relationship between them in order to obtain the 
solution of the entropy inequality. In case of isotropic materials only the 
second order tensors can show cross effects, the vectorial and third order 
tensorial terms are independent. 

The linear relations between the thermodynamic fluxes and forces result in 
the following constitutive equations:
\begin{eqnarray}
m_1 \partial_t q - \partial_x b &=& -l_1 q, \label{ce1}\\
m_2 \partial_t Q - \partial_x B &=& -k_1 Q + k_{12} \partial_x 
q, \label{ce2}\\
b- \frac{1}{T} &=& -k_{21} Q + k_{2} \partial_x q, \label{ce3}\\
B &=& n \partial_xQ. \label{ce4}
\end{eqnarray}
The $l_1, k_1, k_2, n$ coefficients are nonnegative and also 
$K=k_1k_2-k_{12}k_{21} \geq 0$, because of the entropy inequality 
\re{entrpr}. The constitutive equations \re{ce1}-\re{ce4} 
together with the energy balance \re{ebal} and the caloric equation of 
state $e =\rho \hat c T$ form a solvable set of equations. Moreover, in case of 
constant coefficients one can easily eliminate 
the current multipliers by substituting them from \re{ce3}-\re{ce4} into 
\re{ce1}-\re{ce2} and obtain:
\begin{eqnarray}
m_1 \partial_t q + l_1 q - k_2 \partial^2_x q &=& 
	\partial_x \frac{1}{T}- k_{21}\partial_xQ, \label{ce11}\\
m_2 \partial_t Q +k_1 Q - n\partial^2_x Q &=& k_{12} \partial_x q. \label{ce12}
\end{eqnarray}

Here $\partial_x^i$ denotes the $i$-th partial derivative by $x$. 
\re{ce11}-\re{ce12} is identical to \re{qbal}-\re{Qbal} of the previous section 
introducing $\tau_q= m_1/l_1$, $\lambda= (l_1 T^2)^{-1}$, 
$\kappa_{21}=k_{21}/l_1$, $\tau_Q= m_2/k_1$ and $\kappa_{12}=k_{12}/k_1$. If we 
assume that the coefficients  $k_2$ and $n$ are zero we obtain exactly 
\re{qbal}-\re{Qbal}. For the correct sign of the coefficients there one cannot 
assume reciprocity here. Guyer-Krumhansl equation is a special case when $n=0$, 
$m_2=0$ and Maxwell-Cattaneo-Vernotte is obtained if $k_2=0$ and either 
$k_{12}=0$ of $k_{21}=0$ in addition. Several other heat conduction equations 
are obtained in this framework. Jeffreys type is included if $n=0$, $m_1=0$,   
$k_2=0$ and either $k_{12}=0$ of $k_{21}=0$. A Cahn-Hilliard type equation is 
derived if  $n=0$, $m_1=0$ and $m_2=0$ \cite{ForAme08a}. With suitable 
constitutive assumptions several nonlinear equations fit in this general 
framework like the thermomass transport equation \cite{GuoHou10a}, the 
nonlinear  Maxwell-Cattaneo-Vernotte equation derived in \cite{CimEta09a} and 
also the nonlinear Guyer-Krumhansl equation derived in \cite{CimEta10a2}. 

It is remarkable that the Green-Naghdi model III and II cannot be obtained 
here, however it is a valid special case if a general vectorial internal 
variable is introduced instead of the heat flux \cite{VanFul12a}). 

\section{Rational thermomechanics and some other theories}

The common property of this group is the lack of compatibility with kinetic 
theory. The rational approaches are rigorous from mathematical point of view, 
require Noll type material frame indifference and insist to classical entropy 
current density as heat flux divided by temperature.

\subsection{Second viscosity} Both ballistic propagation and second sound can 
be reasonably modeled with the help of second viscosity 
\cite{Rog71a,Ma13a1,Ma13a2}. Second viscosity may have an imaginary part and 
its origin looks like essentially a kind of internal variable theory developed 
on the basis of thermostatics before the advent of non-equilibrium 
thermodynamics. The idea and the method is originally from Mandelstam and 
Leontovitch \cite{ManLeo37a} and is well described in \cite{LanLif59b}.

\subsection{Jeffreys type equation} Jeffreys type equation 
was first suggested by Joseph and Preziosi \cite{JosPre89a} assuming a delayed 
transport in heat conduction. Later on this 
constitutive equation become very popular, because it improves the properties 
of Maxwell-Cattaneo-Vernotte equation from many point of view and because it 
can be obtained from simple (but sometimes unacceptable) assumptions 
\cite{Tzo95a,Zha07b,BriZha09a}. For one dimensional heat pulse experiments the 
Jeffreys type equation is identical with the Guyer-Krumhansl equation.

\subsection{Green-Naghdi equations}
Green and Naghdi derived a particular constitutive equation for the heat flux. 
They have introduced an internal variable, called thermal displacement rate, 
with the particular interpretation being the time derivative of the 
temperature  \cite{GreNag91a}. The corresponding equations coupled to the 
momentum balance are modeling well ballistic propagation and second sound 
\cite{BarSte05a,BarSte07a,BarSte08a,BarFav14a}. A particular case 
of the model of Green and Naghdi is the existence of heat conduction without 
dissipation.  

\subsection{Semi-empirical temperature of Cimmelli and Kos\'inski}  The 
semi-empirical temperature is a scalar internal variable, 
in the framework of a weakly nonlocal theory \cite{CimKos91a,CimAta92a,Cim09a}. 
Coupled to mechanics it reproduces ballistic-wave like-diffusive propagation of 
heat \cite{FriCim98a}. However, in this theory the constitutive theory (both 
the evolution of internal variable and the heat flux -- temperature relation) 
is postulated directly.

\subsection{Rational thermomechanics}

There are many attempts to incorporate Maxwell-Cattaneo-Vernotte type heat 
conduction in the framework of rational thermomechanics 
\cite{JosPre89a,JosPre90a,Str11b}. It is remarkable that a nonlinear version of 
Maxwell-Cattaneo-Vernotte equation seems to fail the test of proper modeling 
the experiment, because of the lack of a certain type of dissipation, 
characteristic  in Guyer-Krumhansl or Jeffreys type models 
\cite{ColNew88a,DreStr93a}. The theory could not meet the challenge obtaining 
compatibility with kinetic theory without the modification of the classical 
entropy current (or the kinetic theory).

\subsection{Other approaches}
There are many more methods of extending thermodynamics beyond local 
equilibrium. The test of these ideas usually starts with heat conduction. One 
may introduce the time derivatives of the state variables as 
additional state variables. However, time derivatives are frame dependent, they 
are many of them and it is not clear which one is to be used. Rigorous 
exploitation methods of the entropy principle do not help in this respect  
\cite{LiuMul72a,Mus90b}. Recently Serdyukov applied this idea for heat 
conduction, too \cite{Ser01a,Ser07a}. The ballistic-diffusive 
equation of Chen is a particular mixture of phenomenological and kinetic 
considerations \cite{Che01a,Che02a}, the thermomass theory of Guo and Hou 
seemingly rediscovers the role of inertia in heat conduction \cite{GuoHou10a}. 

An important class of heat conduction equations is nonlinear by construction 
\cite{ColNew88a,GuoHou10a,CimEta09a,CimEta10a2}. Possible experimental 
verifications require their reliable numerical solution. Analytical results are 
crucial in this respect \cite{Zan99a,Zhu14a}.

These approaches are partially consistent with either kinetic theory or  
continuum mechanics. They may use or avoid the usage of rigorous entropy 
principle exploitation. Some of them, like the delayed differential equations 
of dual phase lag theory \cite{Tzo14b} failed  to fulfill important 
thermodynamic expectations \cite{FabLaz14a,Ruk14a}. None of them were tested 
with experiments of ballistic-wave like-diffusive propagation.

\section{Discussion}

In his mind provoking article Muschik writes that the reason of so many schools 
of thermodynamics is, that one may go beyond local equilibrium with different 
ways. The extension to nonlocal-nonequilibrium is not unique \cite{Musch08p}. 
In this short survey we argued that the different extensions are not equivalent 
regarding their performance of modeling experimental results, neither regarding 
their theoretical consistency and scope. 

The challenge of non-Fouirer heat conduction is to develop a theory that is 
compatible both with kinetic theory and mechanical principles (including 
material frame indifference), passes the tests of existing experiments on 
parallel ballistic, wave like and diffusive propagation of heat and, 
therefore and most importantly, predictive in foretelling new observations and 
phenomena.

This pointed review is far from being complete and the author apologizes of
not mentioning relevant works. This is partially because of his special point 
of view, beyond his limited knowledge.

\section{Acknowledgements}
The work was supported by the grant OTKA K81161 and K104260. Discussion with 
Robert Kov\'acs are appreciated.


\begin{thebibliography}{10}

\bibitem{JosPre89a}
D.~D. Joseph and L.~Preziosi.
\newblock Heat waves.
\newblock {\em Reviews of Modern Physics}, 61:41--73, 1989.

\bibitem{JosPre90a}
D.~D. Joseph and L.~Preziosi.
\newblock Addendum to the paper "heat waves".
\newblock {\em Reviews of Modern Physics}, 62:375--391, 1990.

\bibitem{Cha98a}
DS~Chandrasekharaiah.
\newblock Hyperbolic thermoelasticity: A review of recent literature.
\newblock {\em Applied Mechanics Reviews}, 51(12):705--729, 1998.

\bibitem{Str11b}
B.~Straughan.
\newblock {\em Heat waves}.
\newblock Springer, 2011.

\bibitem{Cim09a}
V.~A. Cimmelli.
\newblock Different thermodynamic theories and different heat conduction laws.
\newblock {\em Journal of Non-Equilibrium Thermodynamics}, 34(4):299--332,
  2009.

\bibitem{Leb14a}
G.~Lebon.
\newblock Heat conduction at micro and nanoscales: A review through the prism
  of extended irreversible thermodynamics.
\newblock {\em Journal of Non-Equilibrium Thermodynamics}, 39(1):35--59, 2014.

\bibitem{Cat48a}
C.~Cattaneo.
\newblock Sulla conduzione del calore.
\newblock {\em Atti Sem. Mat. Fis. Univ. Modena}, 3:83--101, 1948.

\bibitem{JouAta92b}
D.~Jou, J.~Casas-V\'azquez, and G.~Lebon.
\newblock {\em Extended Irreversible Thermodynamics}.
\newblock Springer Verlag, Berlin-etc., 1992.
\newblock 3rd, revised edition, 2001.

\bibitem{MulRug98b}
I.~M\"uller and T.~Ruggeri.
\newblock {\em Rational Extended Thermodynamics}, volume~37 of {\em Springer
  Tracts in Natural Philosophy}.
\newblock Springer Verlag, New York-etc., 2nd edition, 1998.

\bibitem{KosLiu00a}
P.~Kost\"adt and M.~Liu.
\newblock On the causality and stability of the relativistic diffusion
  equation.
\newblock {\em Physical Reviews D}, 62:023003, 2000.

\bibitem{HerPav01a}
L.~Herrera and D.~Pav\'on.
\newblock Why hyperbolic theories of dissipation cannot be ignored: comments on
  a paper by {K}ost\"adt and {L}iu.
\newblock {\em Physical Reviews D}, 64:088503, 2001.

\bibitem{KosLiu01a}
P.~Kost\"adt and Liu M.
\newblock Alleged acausality of the diffusion equations: reply.
\newblock {\em Physical Reviews D}, 64:088504, 2001.

\bibitem{Fic92a}
G.~Fichera.
\newblock Is the {F}ourier theory of heat propagation paradoxical?
\newblock {\em Rediconti del Circolo Matematico di Palermo}, XLI:5--28, 1992.

\bibitem{Cim04a}
V.~A. Cimmelli.
\newblock On the causality requirement for diffusive-hyperbolic systems in
  non-equilibrium thermodynamics.
\newblock {\em Journal of Non-Equilibrium Thermodynamics}, 29(2):125--139,
  2004.

\bibitem{VanBir08a}
P.~V\'an and T.~S. B\'\i{}r\'o.
\newblock Relativistic hydrodynamics - causality and stability.
\newblock {\em The European Physical Journal - Special Topics}, 155:201--212,
  2008.
\newblock arXiv:0704.2039v2.

\bibitem{Pes44a}
V.~Peshkov.
\newblock Second sound in {H}elium {II}.
\newblock {\em J. Phys. (Moscow)}, 8:381, 1944.

\bibitem{Tisz38a}
L.~Tisza.
\newblock Transport phenomena in helium {II}.
\newblock {\em Nature}, 141:913, 1938.

\bibitem{Lan41a}
L.D. Landau.
\newblock Two-fluid model of liquid helium {II}.
\newblock {\em Journal of Physics}, 5(1):71--90, 1941.

\bibitem{MorFes53b}
P.~M. Morse and H.~Feshbach.
\newblock {\em Methods of theoretical physics}.
\newblock McGraw-Hill, 1953.

\bibitem{Ver58a}
M.~P. Vernotte.
\newblock La v\'eritable \'equation de chaleur.
\newblock {\em Comptes rendus hebdomadaires des s\'eances de l'Acad\'emie des
  sciences}, 247:2103--5, 1958.

\bibitem{GuyKru64a}
R.~A. Guyer and J.~A. Krumhansl.
\newblock Dispersion relation for a second sound in solids.
\newblock {\em Physical Review}, 133:A1411, 1964.

\bibitem{GuyKru66a1}
R.~A. Guyer and J.~A. Krumhansl.
\newblock Solution of the linearized phonon {B}oltzmann equation.
\newblock {\em Physical Review}, 148(2):766--778, 1966.

\bibitem{GuyKru66a2}
R.~A. Guyer and J.~A. Krumhansl.
\newblock Thermal conductivity, second sound and phonon hydrodynamic phenomena
  in nonmetallic crystals.
\newblock {\em Physical Review}, 148(2):778--788, 1966.

\bibitem{AckGuy68a}
C.C. Ackerman and R.A. Guyer.
\newblock Temperature pulses in dielectric solids.
\newblock {\em Annals of Physics}, 50(1):128--185, 1968.

\bibitem{AckOve69a}
C.C. Ackerman and W.C. Overton.
\newblock Second sound in solid helium-3.
\newblock {\em Physical Review Letters}, 22(15):764, 1969.

\bibitem{McNEta70a}
T.F. McNelly, S.J. Rogers, D.J. Channin, R.J. Rollefson, W.M. Goubau, G.E.
  Schmidt, J.A. Krumhansl, and R.O. Pohl.
\newblock Heat pulses in {NaF}: onset of second sound.
\newblock {\em Physical Review Letters}, 24(3):100, 1970.

\bibitem{NarDyn72a}
V.~Narayanamurti and R.~D. Dynes.
\newblock Observation of second sound in {B}ismuth.
\newblock {\em Physical Review Letters}, 26:1461--1465, 1972.

\bibitem{JacWal71a}
H.~E. Jackson and C.~T. Walker.
\newblock Thermal conductivity, second sound and phonon-phonon interactions in
  {N}a{F}.
\newblock {\em Physical Review B}, 3(4):1428--1439, 1971.

\bibitem{Kam90a}
W.~Kaminski.
\newblock Hyperbolic heat conduction equations for materials with a
  nonhomogeneous inner structure.
\newblock {\em Journal of Heat Transfer}, 112:555--560, 1990.

\bibitem{MitEta95a}
K.~Mitra, S.~Kumar, A.~Vedavarz, and Moallemi~M. K.
\newblock Experimental evidence of hyperbolic heat conduction in processed
  meat.
\newblock {\em Journal of Heat Transfer}, 117:568--573, 1995.

\bibitem{HerBec00a}
H.~Herwig and K.~Beckert.
\newblock Fourier versus non-{F}ourier heat conduction in materials with a
  nonhomogeneous inner structure.
\newblock {\em Journal of Heat Transfer}, 122(2):363--365, 2000.

\bibitem{HerBec00a1}
H.~Herwig and K.~Beckert.
\newblock Experimental evidence about the controversy concerning {F}ourier or
  non-{F}ourier heat conduction in materials with a nonhomogeneous inner
  structure.
\newblock {\em Heat and Mass Transfer}, 36(5):387--392, 2000.

\bibitem{RoeEta03a}
W.~Roetzel, N.~Putra, and S.~K. Das.
\newblock Experiment and analysis for non-{F}ourier conduction in materials
  with non-homogeneous inner structure.
\newblock {\em International Journal of Thermal Sciences}, 42(6):541--552,
  2003.

\bibitem{ScoEta09a}
E.~P. Scott, M.~Tilahun, and B.~Vick.
\newblock The question of thermal waves in heterogeneous and biological
  materials.
\newblock {\em Journal of Biomechanical Engineering}, 131:074518, 2009.

\bibitem{DasEta03a}
S.~K. Das, N.~Putra, P.~Thiesen, and W.~Roetzel.
\newblock Temperature dependence of thermal conductivity enhancement for
  nanofluids.
\newblock {\em Journal of Heat Transfer}, 125(4):567--574, 2003.

\bibitem{KimEta01a}
P.~Kim, L.~Shi, A.~Majumdar, and P.~L. Mc{E}uen.
\newblock Thermal transport measurements of individual multiwalled nanotubes.
\newblock {\em Physical Review Letters}, 87(21):215502, 2001.

\bibitem{CahEta03a}
D.G. Cahill, W.~K. Ford, K.~E. Goodson, G.~D. Mahan, A.~Majumdar, H.~J. Maris,
  R.~Merlin, and S.~R. Phillpot.
\newblock Nanoscale thermal transport.
\newblock {\em Journal of Applied Physics}, 93(2):793--818, 2003.

\bibitem{Zha07b}
Z.~M. Zhang.
\newblock {\em Nano/microscale heat transfer}.
\newblock McGrawHill, New York, etc..., 2007.

\bibitem{CimEta09a}
V.~A. Cimmelli, A.~Sellitto, and D.~Jou.
\newblock Nonlocal effects and second sound in a nonequilibrium steady state.
\newblock {\em Physical Review B}, 79:014303, 2009.

\bibitem{CimEta13a}
A.~Sellitto, V.~A. Cimmelli, and D.~Jou.
\newblock Entropy flux and anomalous axial heat transport at the nanoscale.
\newblock {\em Physical Review B}, 87:054302(7), 2013.

\bibitem{LuzRom08a}
M.~Luzum and P.~Romatschke.
\newblock Conformal relativistic viscous hydrodynamics: Applications to rhic
  results at $\sqrt{s_{NN}}$ = 200 {GeV}.
\newblock {\em Physical Review C}, 78(3):034915, 2008.

\bibitem{Mul67a}
I.~M\"uller.
\newblock On the entropy inequality.
\newblock {\em Archive for Rational Mechanics and Analysis}, 26(2):118--141,
  1967.

\bibitem{Isr76a}
W.~Israel.
\newblock Nonstationary irreversible thermodynamics - causal relativistic
  theory.
\newblock {\em Annals of Physics}, 100(1-2):310--331, 1976.

\bibitem{IsrSte76a}
W.~Israel and J.~M. Stewart.
\newblock Thermodynamics of nonstationary and transient effect in a
  relativistic gas.
\newblock {\em Physics Letters A}, 58(4):213--215, 1976.

\bibitem{VanBir12a}
P.~V\'an and T.S. Bir\'o.
\newblock First order and generic stable relativistic dissipative
  hydrodynamics.
\newblock {\em Physics Letters B}, 709(1-2):106--110, 2012.
\newblock arXiv:1109.0985[nucl-th].

\bibitem{CimEta14a}
V.~A. Cimmelli, D.~Jou, T.~Ruggeri, and P.~V\'an.
\newblock Entropy principle and recent results in non-equilibrium theories.
\newblock {\em Entropy}, 16:1756--1807, 2014.
\newblock http://www.mdpi.com/1099-4300/16/3/1756.

\bibitem{LebEta08b}
G.~Lebon, D.~Jou, and J.~Casas-V{\'a}zquez.
\newblock {\em Understanding non-equilibrium thermodynamics}.
\newblock Springer, 2008.

\bibitem{LebAta98a}
G.~Lebon, D.~Jou, J.~Casas-V\'azquez, and W.~Muschik.
\newblock Weakly nonlocal and nonlinear heat transport in rigid solids.
\newblock {\em Journal of Non-Equilibrium Thermodynamics}, 23:176--191, 1998.

\bibitem{DreStr93a}
W.~Dreyer and H.~Struchtrup.
\newblock Heat pulse experiments revisited.
\newblock {\em Continuum Mechanics and Thermodynamics}, 5:3--50, 1993.

\bibitem{Eck40a1}
C.~Eckart.
\newblock The thermodynamics of irreversible processes, {I}. {T}he simple
  fluid.
\newblock {\em Physical Review}, 58:267--269, 1940.

\bibitem{Eck40a2}
C.~Eckart.
\newblock The thermodynamics of irreversible processes, {II}. {F}luid mixtures.
\newblock {\em Physical Review}, 58:269--275, 1940.

\bibitem{Eck40a3}
C.~Eckart.
\newblock The thermodynamics of irreversible processes, {III}. {R}elativistic
  theory of the simple fluid.
\newblock {\em Physical Review}, 58:919--924, 1940.

\bibitem{Eck48a}
C.~Eckart.
\newblock The thermodynamics of irreversible processes. {IV}. {T}he theory of
  elasticity and anelasticity.
\newblock {\em Physical Review}, 73(4):373--382, 1948.

\bibitem{Mul67a1}
I.~M\"uller.
\newblock Zur paradoxon der {W}\"armeleitungstheorie.
\newblock {\em Zeitschrift f\"ur Physik}, 198:329--344, 1967.

\bibitem{MacOns53a}
S.~Machlup and L.~Onsager.
\newblock Fluctuations and irreversible processes. {II}. {S}ystems with kinetic
  energy.
\newblock {\em Physical Review}, 91(6):1512--1515, 1953.

\bibitem{Gya77a}
I.~Gyarmati.
\newblock The wave approach of thermodynamics and some problems of non-linear
  theories.
\newblock {\em Journal of Non-Equilibrium Thermodynamics}, 2:233--260, 1977.

\bibitem{Ver85b}
Verh\'as J.
\newblock {\em Termodinamika \'es reol\'ogia}.
\newblock M\H{u}szaki K\"onyvkiad\'o, Budapest, 1985.

\bibitem{Ver97b}
J.~Verh\'as.
\newblock {\em Thermodynamics and {R}heology}.
\newblock Akad\'emiai Kiad\'o and Kluwer Academic Publisher, Budapest, 1997.

\bibitem{Ver83a}
J.~Verh\'as.
\newblock On the entropy current.
\newblock {\em Journal of Non-Equilibrium Thermodynamics}, 8:201--206, 1983.

\bibitem{CimVan05a}
V.~A. Cimmelli and P.~V\'an.
\newblock The effects of nonlocality on the evolution of higher order fluxes in
  non-equilibrium thermodynamics.
\newblock {\em Journal of Mathematical Physics}, 46(11):112901--15, 2005.
\newblock cond-mat/0409254.

\bibitem{CiaAta07a}
V.~Ciancio, V.~A. Cimmelli, and P.~V\'an.
\newblock On the evolution of higher order fluxes in non-equilibrium
  thermodynamics.
\newblock {\em Mathematical and Computer Modelling}, 45:126--136, 2007.
\newblock cond-mat/0407530.

\bibitem{Nyi91a1}
B.~Ny\'\i{}ri.
\newblock On the entropy current.
\newblock {\em Journal of Non-Equilibrium Thermodynamics}, 16:179--186, 1991.

\bibitem{Van01a2}
P.~V\'an.
\newblock Weakly nonlocal irreversible thermodynamics -- the
  {G}uyer-{K}rumhansl and the {C}ahn-{H}illiard equations.
\newblock {\em Physics Letters A}, 290(1-2):88--92, 2001.
\newblock (cond-mat/0106568).

\bibitem{KovVan15a}
R.~Kov\'acs and P.~V\'an.
\newblock Generalized heat conduction in heat pulse experiments.
\newblock {\em International Journal of Heat and Mass Transfer}, 83:613--620,
  2015.
\newblock arXiv:1409.0313v2.

\bibitem{ForAme08a}
S.~Forest and M.~Amestoy.
\newblock Hypertemperature in thermoelastic solids.
\newblock {\em Comptes Rendus Mecanique}, 336:347--353, 2008.

\bibitem{GuoHou10a}
Zeng-Yuan Guo and Quan-Wen Hou.
\newblock Thermal wave based on the thermomass model.
\newblock {\em Journal of Heat Transfer}, 132(7):072403, 2010.

\bibitem{CimEta10a2}
V.~A. Cimmelli, A.~Sellitto, and D.~Jou.
\newblock Nonlinear evolution and stability of the heat flow in nanosystems:
  {B}eyond linear phonon hydrodynamics.
\newblock {\em Physical Review B}, 82:184302, 2010.

\bibitem{VanFul12a}
P.~V\'an and T.~F\"ul\"op.
\newblock Universality in heat conduction theory: weakly nonlocal
  thermodynamics.
\newblock {\em Annalen der Physik}, 524(8):470--478, 2012.
\newblock arXiv:1108.5589.

\bibitem{Rog71a}
SJ~Rogers.
\newblock Transport of heat and approach to second sound in some isotopically
  pure alkali-halide crystals.
\newblock {\em Physical Review B}, 3(4):1440, 1971.

\bibitem{Ma13a1}
Yanbao Ma.
\newblock A transient ballistic--diffusive heat conduction model for heat pulse
  propagation in nonmetallic crystals.
\newblock {\em International Journal of Heat and Mass Transfer}, 66:592--602,
  2013.

\bibitem{Ma13a2}
Yanbao Ma.
\newblock A hybrid phonon gas model for transient ballistic-diffusive heat
  transport.
\newblock {\em Journal of Heat Transfer}, 135(4):044501, 2013.

\bibitem{ManLeo37a}
L.I. Mandelstam and M.A. Leontovich.
\newblock On sound dissipation theory in fluids.
\newblock {\em JETP}, 7(3):438--448, 1937.

\bibitem{LanLif59b}
L.~D. Landau and E.~M. Lifshitz.
\newblock {\em Fluid mechanics}.
\newblock Pergamon Press, London, 1959.

\bibitem{Tzo95a}
Da~Yu Tzou.
\newblock The generalized lagging response in small-scale and high-rate
  heating.
\newblock {\em J. Heat Mass Transfer}, 38(17):3231--3240, 1995.

\bibitem{BriZha09a}
T.~J. Bright and Z.~M. Zhang.
\newblock Common misperceptions of the hyperbolic heat equation.
\newblock {\em Journal of Thermophysics and Heat Transfer}, 23(3):601--607,
  2009.

\bibitem{GreNag91a}
A.E. Green and P.~M. Naghdi.
\newblock A re-examination of the basic postulates of thermomechanics.
\newblock {\em Proceedings of the Royal Society: Mathematical and Physical
  Sciences}, 432(1885):171--194, 1991.

\bibitem{BarSte05a}
S.~Bargmann and P.~Steinmann.
\newblock Finite element approaches to non-classical heat conduction in solids.
\newblock {\em Computer Modeling in Engineering Sciences}, 9(2):133–150,
  2005.

\bibitem{BarSte07a}
S.~Bargmann and P.~Steinmann.
\newblock Classical results for a non-classical theory: remarks on
  thermodynamic relations in {G}reen-–{N}aghdi thermo-hyperelasticity.
\newblock {\em Continuum Mechanics and Thermodynamics}, 19:59--66, 2007.

\bibitem{BarSte08a}
S.~Bargmann and P.~Steinmann.
\newblock Modeling and simulation of first and second sound in solids.
\newblock {\em International Journal of Solids and Structures}, 45:6067--6073,
  2008.

\bibitem{BarFav14a}
S.~Bargmann and A.~Favata.
\newblock Continuum mechanical modeling of laser-pulsed heating in
  polycrystals: {A} multi-physics problem of coupling diffusion, mechanics, and
  thermal waves.
\newblock {\em ZAMM-Journal of Applied Mathematics and Mechanics/Zeitschrift
  f{\"u}r Angewandte Mathematik und Mechanik}, 94(6):487--498, 2014.

\bibitem{CimKos91a}
V.A. Cimmelli and W.~Kosinski.
\newblock Nonequilibrium semi-empirical temperature in materials with thermal
  relaxation.
\newblock {\em Archives of Mechanics}, 43(6):753--767, 1991.

\bibitem{CimAta92a}
V.~A. Cimmelli, W.~Kosi\'nski, and K.~Saxton.
\newblock Modified {F}ourier law -- comparison of two approaches.
\newblock {\em Archive of Mechanics}, 44(4):409--415, 1992.

\bibitem{FriCim98a}
K.~Frischmuth and V.A. Cimmelli.
\newblock Coupling in thermo-mechanical wave propagation in {NaF} at low
  temperature.
\newblock {\em Archives of Mechanics}, 50:703--714, 1998.

\bibitem{ColNew88a}
B.~D. Coleman and D.~Newman.
\newblock Implications of a nonlinearity in the theory of second sound in
  solids.
\newblock {\em Physical Review B}, 37(4):1492, 1988.

\bibitem{LiuMul72a}
I-Shih Liu and I.~M\"uller.
\newblock On the thermodynamics and thermostatics of fluids in electromagnetic
  fields.
\newblock {\em Archive for Rational Mechanics and Analysis}, 46(2):149--176,
  1972.

\bibitem{Mus90b}
W.~Muschik.
\newblock {\em Aspects of Non-Equilibrium Thermodynamics}.
\newblock World Scientific, Singapure-etc., 1990.

\bibitem{Ser01a}
S.~I. Serdyukov.
\newblock A new version of extended irreversible thermodynamics and
  dual-phase-lag model in heat transfer.
\newblock {\em Physics Letters A}, 281:16--20, 2001.

\bibitem{Ser07a}
S.~I. Serdyukov.
\newblock On the definitions of entropy and temperature in the extended
  thermodynamics of irreversible processes.
\newblock {\em C. R. Physique}, 8:93--100, 2007.

\bibitem{Che01a}
G.~Chen.
\newblock Ballistic-diffusive heat-conduction equations.
\newblock {\em Physical Review Letters}, 86(11):2297(4), 2001.

\bibitem{Che02a}
G.~Chen.
\newblock Ballistic-diffusive equations for transient heat conduction from nano
  to macroscales.
\newblock {\em ASME Journal of Heat Transfer}, 124(2):320--328, 2002.

\bibitem{Zan99a}
E.~Zanchini.
\newblock Hyperbolic-heat-conduction theories and nondecreasing entropy.
\newblock {\em Physical Review B}, 60(2):991, 1999.

\bibitem{Zhu14a}
K.~Zhukovsky.
\newblock Solution of some types of differential equations: Operational
  calculus and inverse differential operators.
\newblock {\em The Scientific World Journal}, 2014:454865, 2014.

\bibitem{Tzo14b}
D.~Y. Tzou.
\newblock {\em Macro- to microscale heat transfer ({T}he lagging behaviour)}.
\newblock Taylor and Francis, 2 edition, 2014.

\bibitem{FabLaz14a}
M.~Fabrizio and B.~Lazzari.
\newblock Stability and second law of thermodynamics in dual-phase-lag heat
  conduction.
\newblock {\em International Journal of Heat and Mass Transfer}, 74:484--489,
  2014.

\bibitem{Ruk14a}
Sergey~A Rukolaine.
\newblock Unphysical effects of the dual-phase-lag model of heat conduction.
\newblock {\em International Journal of Heat and Mass Transfer}, 78:58--63,
  2014.

\bibitem{Musch08p}
W.~Muschik.
\newblock Why so many “schools” of thermodynamics?
\newblock {\em Atti dell'Accademia Peloritana dei Pericolanti, Classe di
  Scienze Fisiche, Matematiche e Naturali, Suppl. I.}, 86:C1S0801002, 2008.

\end{thebibliography}

\end{document}